\documentclass[aps,prd,preprintnumbers,groupedaddress,nofootinbib,amssymb,eqsecnum,notitlepage]{revtex4}
\usepackage{graphicx}
\usepackage{bm}
\usepackage{amsmath}
\usepackage{color}
\usepackage{amsfonts}
\usepackage{here}
\usepackage{graphicx}
\usepackage{amsmath,amsthm,amssymb}
\usepackage{bm}
\allowdisplaybreaks[1]

\usepackage{amsfonts}
\usepackage{dcolumn}
\usepackage{hyperref}

\begin{document}
\newcommand{\newc}{\newcommand}

\newcommand{\ben}{\begin{eqnarray}}
\newcommand{\een}{\end{eqnarray}}
\newc{\be}{\begin{equation}}
\newc{\ee}{\end{equation}}
\newc{\ba}{\begin{eqnarray}}
\newc{\ea}{\end{eqnarray}}
\newc{\bea}{\begin{eqnarray*}}
\newc{\eea}{\end{eqnarray*}}
\newc{\da}{\delta{A}}
\newc{\D}{\partial}
\newc{\ie}{{\it i.e.} }
\newc{\eg}{{\it e.g.} }
\newc{\etc}{{\it etc.} }
\newc{\etal}{{\it et al.}}
\newcommand{\nn}{\nonumber}
\newc{\ra}{\rightarrow}
\newc{\lra}{\leftrightarrow}
\newc{\lsim}{\buildrel{<}\over{\sim}}
\newc{\gsim}{\buildrel{>}\over{\sim}}
\newc{\aP}{\alpha_{\rm P}}
\newc{\rk}[1]{\textcolor{red}{#1}}
\newc{\mm}[1]{\textcolor{red}{#1}}
\newcommand{\mmc}[1]{\textcolor{green}{[MM:~#1]}}
\newc{\Mpl}{M_{\rm pl}}

\title{Black holes in quartic-order 
beyond-generalized Proca theories}

\author{
Ryotaro Kase$^{1}$,
Masato Minamitsuji$^{2}$, and
Shinji Tsujikawa$^{1}$
}

\affiliation{
$^1$Department of Physics, Faculty of Science, Tokyo University of Science, 1-3, Kagurazaka,
Shinjuku-ku, Tokyo 162-8601, Japan\\
$^2$Centro de Astrof\'{\i}sica e Gravita\c c\~ao  - CENTRA,
Departamento de F\'{\i}sica, Instituto Superior T\'ecnico - IST,
Universidade de Lisboa - UL, Av. Rovisco Pais 1, 1049-001 Lisboa, Portugal.
}

\date{\today}

\begin{abstract}

The generalized Proca theories with second-order equations 
of motion can be healthily extended to a more general framework 
in which the number of 
propagating degrees of freedom remains unchanged.
In the presence of a quartic-order nonminimal coupling to gravity 
arising in beyond-generalized Proca theories, the speed of gravitational 
waves $c_t$ on the Friedmann-Lema\^{i}tre-Robertson-Walker (FLRW) 
cosmological background can be equal to 
that of light $c$ under a certain condition. 
By using this condition alone, we show that the speed of 
gravitational waves in the vicinity of static and spherically 
symmetric black holes is also equivalent to $c$ for the propagation 
of odd-parity perturbations along both radial 
and angular directions. 
As a by-product, the black holes arising in our 
beyond-generalized Proca theories 
are plagued by neither ghost nor Laplacian instabilities 
against odd-parity perturbations. 
We show the existence of both exact and numerical 
black hole solutions endowed with vector hairs induced 
by the quartic-order coupling.

\end{abstract}


\maketitle

\section{Introduction}
\label{introsec}

The constantly accumulating observational evidence of dark energy 
and dark matter implies the existence of additional degrees of freedom (DOFs) beyond those appearing in standard model 
of physics or General Relativity (GR) \cite{review}.
One of the candidates for such extra DOFs is a spin-0 scalar 
field $\phi$. If the scalar field is nonminimally coupled 
to gravity, Horndeski theories \cite{Horndeski} 
are the most general scalar-tensor theories with second-order 
equations of motion \cite{redis}. 
It is also possible to perform a healthy extension of 
Horndeski theories without increasing the propagating 
DOFs (one scalar and two tensor polarizations) \cite{GLPV,Langlois,Cris}.

The other candidate for extra DOFs is a spin-1 vector field 
$A_{\mu}$. A massless vector field respects the $U(1)$ gauge 
symmetry in Minkowski spacetime, but the gauge invariance 
is explicitly broken by introducing a vector-field mass or 
by considering derivative and nonminimal couplings. 
Most general $U(1)$-broken vector-tensor theories with 
second-order equations of motion are known as generalized 
Proca (GP) theories \cite{Heisenberg,Tasinato,Allys,Jimenez}, 
which contain five propagating DOFs 
(one longitudinal scalar, two transverse vectors, 
and two tensor polarizations).
If we apply GP theories to cosmology, there exists an 
interesting de Sitter attractor responsible for the late-time 
cosmic acceleration \cite{GPcosmo1}. 
The dark energy models in the framework of 
GP theories are observationally distinguished from the 
cosmological constant due to different cosmic expansion and 
growth histories \cite{GPcosmo2,GPcosmo3}. 
One can extend GP theories to the domain of beyond-generalized 
Proca (BGP) theories \cite{HKT16,KNY16,Naka} 
in which the propagating DOFs remain five.

The recent gravitational-wave (GW) event GW170817 \cite{GW170817} 
from a neutron star merger, together with the gamma-ray 
burst GRB 170817A \cite{Goldstein},  
showed that the speed of gravitational waves $c_t$ 
traveling over a cosmological distance (the redshift $z<0.009$) 
is very close to that of light $c$ 
with the difference less than the order of $10^{-15}$.
If we demand that $c_t$ is strictly equivalent to $c$, 
neither quartic-order nor quintic-order nonminimal 
derivative couplings appearing in Horndeski 
and GP theories are 
allowed \cite{GW1,GW2,GW3,GW4} 
(see also Refs.~\cite{GW5,GW6}).
In scalar-tensor theories beyond Horndeski, it is possible 
to realize $c_t=c$ on the FLRW cosmological background 
even in the presence of quartic-order nonminimal 
derivative couplings \cite{building,KaseIJ}. 
This is also the case for 
quartic-order BGP theories \cite{HKT16}.

After the detection of GWs from a black hole (BH) 
merger \cite{Abbott}, we are 
now entering an era in which the physics of BHs can be probed from precise GW measurements in nonlinear regimes of gravity.
In theories beyond GR, the existence of extra DOFs can leave 
imprints on BH solutions 
as new ``hairs''. 
In Horndeski theories, for example, there are several hairy BH 
solutions on a static and spherically symmetric background 
for a radial-dependent scalar 
$\phi=\phi(r)$ \cite{Rinaldi,Anabalon,Minami13,Soti1,Soti2,GB1,GB2,GB3,GB4} 
or a linearly time-dependent scalar 
$\phi=qt+\psi(r)$ \cite{Babi14,Koba14} 
(see also Ref.~\cite{Babi16_2} and references therein).
In the latter configuration there exists a stealth 
Schwarzschild solution for the quartic coupling $G_4$ containing 
a linear term of $\partial_{\mu}\phi \partial^{\mu}\phi$
and the reduced Planck mass squared $M_{\rm pl}^2$, in which case 
the GW speed differs from $c$ on the cosmological background.
The quartic-order beyond-Horndeski interaction allows for 
the realization of a 
model in which $c_t$ is equivalent to $c$ \cite{Babi18}.

In GP theories, the existence of a temporal vector component $A_0$ besides a longitudinal component $A_1$ gives rise to a wide variety of hairy BH solutions \cite{Chagoya,Fan,Minami16,Cisterna,Chagoya2,Babichev17,HKMT,HKMT2,Chagoya18,Filippini}. 
For example, there is a stealth Schwarzschild solution with 
$A_1 \neq 0$ for the specific quartic coupling 
$G_4(X)=M_{\rm pl}^2/2+X/4$,
where $X=-A_{\mu}A^{\mu}/2$. 
Recently, it was shown that this BH solution is unstable 
against odd-parity perturbations in the vicinity of 
the event horizon \cite{KMTZ}. 
The point is that, under the absence of ghosts, 
one of the propagation speed squares 
along the angular direction is negative.
This instability problem is intrinsically related to the 
fact that the speed of GWs around BHs 
is different from $c$ for quartic couplings $G_4(X)$.  
There is also the branch with $A_1=0$, but the model 
given by the coupling 
$G_4(X)=M_{\rm pl}^2/2+\beta_4 M_{\rm pl}^2 
(X/M_{\rm pl}^2)^n$ with $n \geq 1$ also 
leads to the radial and angular propagation speeds 
whose deviations from $c$ approach nonvanishing 
constants at spatial infinity \cite{KMTZ}. 
Unless the coupling $\beta_4$ is very small, 
this behavior is at odds with the observed 
speed of GWs. 
The extension to BGP theories can give rise to 
the exact value $c_t=c$, so there is a possibility for 
overcoming the above mentioned problems.

In this paper, we focus on quartic-order BGP theories 
and study whether the condition imposed for obtaining 
the value $c_t=c$ on the FLRW cosmological background 
is sufficient for realizing the same speed 
of GWs in the vicinity of BHs. 
In Sec.~\ref{modelsec}, we derive the equations 
of motion in quartic-order BGP theories on 
a static and spherically symmetric background. 
In Sec.~\ref{stasec}, we obtain the propagation speeds 
of GW and vector-field perturbation in the vicinity of 
BHs by considering odd-parity perturbations. 
We show that the condition for realizing
the cosmological value $c_t=c$ is sufficient to 
obtain the same propagation speed around BHs. 
In Sec.~\ref{BHsec}, we search for exact 
and numerical BH solutions with vector hairs 
in BGP theories satisfying 
$c_t=c$. 
As a result, our new hairy BH solutions 
are affected by neither ghost nor Laplacian instabilities 
against odd-parity perturbations. 
In the rest of sections, we choose the natural unit $c=1$.

\section{Quartic-order beyond-generalized Proca theories}
\label{modelsec}

We consider quartic-order BGP theories \cite{HKT16}
with the vector field $A_{\mu}$ and the field strength
$F_{\mu \nu}=\nabla_{\mu}A_{\nu}-\nabla_{\nu}A_{\mu}$, 
where $\nabla_\mu$ is the covariant derivative 
operator.
The corresponding action is given by 
\be
S=\int d^{4}x \sqrt{-g} 
\left[ -\frac{1}{4}F_{\mu \nu}F^{\mu \nu}+ G_{4}(X) R 
+ G_{4,X}(X)\left\{ (\nabla_{\mu} A^{\mu})^{2} 
-  \nabla_{\mu} A_{\nu} \nabla^{\nu} A^{\mu}\right\} 
+{\cal L}_4^{\rm BGP}
\right]\,,
\label{action}
\ee
where $g$ is the determinant of four-dimensional 
metric tensor $g_{\mu\nu}$, $R$ is the Ricci scalar,
and $G_4$ is a function of $X= -A_{\mu}A^{\mu}/2$ 
with the notation 
$G_{4,X} \equiv \partial G_4/\partial X$.
The Lagrangian ${\cal L}_4^{\rm BGP}$ is a new term 
appearing beyond the domain of second-order 
GP theories, which is given by 
\be
{\cal L}_4^{\rm BGP}=f_4(X) 
{\cal E}_{\alpha_1 \alpha_2\alpha_3\gamma_4}
{\cal E}^{\beta_1 \beta_2\beta_3\gamma_4}
A^{\alpha_1}A_{\beta_1}
\nabla^{\alpha_2}A_{\beta_2} 
\nabla^{\alpha_3}A_{\beta_3}\,,
\label{L4BGP}
\ee
where $f_4$ is a function of $X$, and
${\cal E}_{\alpha_1 \alpha_2\gamma_3\gamma_4}$ 
is the Levi-Civita tensor satisfying the normalization  
${\cal E}_{\alpha_1 \alpha_2\gamma_3\gamma_4}{\cal E}^{\alpha_1 \alpha_2\gamma_3\gamma_4}=-4!$. 
We note that, by taking the scalar limit $A_{\mu}\to\nabla_{\mu}\phi$, 
the action (\ref{action}) reduces to that of quartic-order shift-symmetric 
Horndeski theories and its Gleyzes-Langlois-Piazza-Vernizzi (GLPV) 
extension \cite{GLPV}.

We study BH solutions on a static and spherically
symmetric background described by the line element
\be
ds^{2} =-f(r) dt^{2} +h^{-1}(r)dr^{2} + 
r^{2} (d\theta^{2}+\sin^{2}\theta\, d \varphi^{2}),
\label{metric}
\ee
where $t$, $r$ and $(\theta,\varphi)$ represent the time, 
radial, and angular coordinates, respectively, 
and $f, h$ are functions of $r$.
The vector-field profile compatible with 
the background (\ref{metric}) is \cite{screening}
\be
A_{\mu}=\left( A_0 (r), A_1(r), 0, 0 \right)\,,
\label{Amu}
\ee
where $A_0$ and $A_1$ are functions of $r$.
The quantity $X$ is expressed in the form 
\be
X=\frac{A_0^2}{2f}-\frac{hA_1^2}{2}\,. 
\ee
We compute the action (\ref{action})
on the background (\ref{metric}) and vary
it with respect to $f, h, A_0, A_1$.
The resulting equations of motion are
given by 
\ba
& &
\frac{c_{1}}{r}h' 
+ c_{2} + \frac{c_{3}}{r} + \frac{c_{4}}{r^{2}}
=0\,, 
\label{be1} \\
& &
-\frac{h}{f} \frac{c_{1}}{r}f' 
+ c_{5} + \frac{c_{6}}{r} + \frac{c_{7}}{r^{2}}
=0\,,
\label{be2}\\
& & \left( d_1+\frac{d_2}{r}\right)f'
+\left( d_3+\frac{d_4}{r} \right)h'
+d_5+\frac{d_6}{r}+\frac{d_7}{r^2}=0\,,
\label{be3}\\
& & d_{8}f'
+d_{9}+\frac{d_{10}}{r}=0\,,
\label{be4}
\ea
where a prime represents the derivative 
with respect to $r$.
The coefficients $c_{1},\cdots, c_{7}$ and 
$d_{1},\cdots, d_{10}$ are given in Appendix~\ref{appendix1}.

On the FLRW cosmological background, 
the propagation speed $c_t$ of tensor perturbations 
was computed in Ref.~\cite{HKT16}.
For the theories given by the action 
(\ref{action}), we have
\be
c_t^2=\frac{G_4}{G_4-2XG_{4,X}-4X^2 f_4}\,.
\ee
The condition for realizing the value $c_t^2=1$ translates to 
\be
f_4=-\frac{G_{4,X}}{2X}\,,
\label{f4con}
\ee
where $X \neq 0$. 
In Sec.~\ref{stasec}, we show that, under the condition 
(\ref{f4con}), the propagation speed squared of gravitational 
waves in the odd-parity sector around the static and spherically 
symmetric background (\ref{metric}) 
is also equivalent to 1.
In Sec.~\ref{BHsec}, we search for hairy BH solutions 
by imposing the condition (\ref{f4con}).

\section{Odd-parity perturbations}
\label{stasec}

We study the stability of BHs against odd-parity perturbations 
on top of the spacetime metric (\ref{metric}) and 
the vector-field profile (\ref{Amu}).
We decompose the metric $g_{\mu \nu}$ and the vector field $A_{\mu}$ 
into the background and perturbed parts as 
$g_{\mu \nu}=\bar{g}_{\mu \nu}+h_{\mu \nu}$ and 
$A_{\mu}=\bar{A}_{\mu}+\delta A_{\mu}$, where a 
bar represents the background values. 
The components of metric perturbations $h_{\mu \nu}$ 
in the odd-parity sector are expressed in the 
forms \cite{GB,Motohashi,KMTZ,Kobayashi1,Kobayashi2,Taka}:
\ba
&&
h_{tt}=h_{tr}=h_{rr}=0\,,\label{htt} \\
&&
h_{ta}=\sum_{l,m}Q_{lm}(t,r)E_{ab}\partial^bY_{lm}(\theta,\varphi)\,,
\label{Qlm}\\
& &
h_{ra}=\sum_{l,m}W_{lm}(t,r)E_{ab}\partial^bY_{lm}(\theta,\varphi)\,,\\
&&
h_{ab}=\frac{1}{2}\sum_{lm}
U_{lm} (t,r)
\left[
E_{a}{}^c \nabla_c\nabla_b Y_{lm}(\theta,\varphi)
+ E_{b}{}^c \nabla_c\nabla_a Y_{lm}(\theta,\varphi)
\right]\,,
\label{Ulm}
\ea
where $a, b$ represent $\theta$ or $\varphi$, and 
$Q_{lm}$, $W_{lm}$, $U_{lm}$ are functions 
of $t$ and $r$. 
The tensor $E_{ab}$ is defined by 
$E_{ab}=\sqrt{\gamma}\, \varepsilon_{ab}$, where 
$\gamma$ is the determinant of two-dimensional metric 
$\gamma_{ab}$ on the sphere
and $\epsilon_{ab}$ is the Levi-Civita symbol
with $\epsilon_{\theta\varphi}=1$,
and $Y_{lm}$ is the 
spherical harmonics.
We choose the Regge-Wheller gauge 
\cite{Regge:1957td,Zerilli:1970se}, 
in which the perturbation $U_{lm}$ vanishes.
The vector perturbation $\delta A_{lm}$ for 
odd-parity modes is given by 
\be
\da_{t}=\da_{r}=0\,,\qquad
\da_{a}=\sum_{l,m}\da_{lm}(t,r)E_{ab}\partial^bY_{lm}
(\theta,\varphi)\,,
\label{vectorper}
\ee
where $\da_{lm}$ is a function of $t$ and $r$. 

We expand the action (\ref{action}) up to quadratic 
order in odd-parity perturbations and then perform 
the integrals with respect to $\theta$ and $\varphi$.
Integrating the action by parts with respect to $t$ and $r$, 
and using the background equations of motion 
(\ref{be1})-(\ref{be4}),  
we obtain the second-order action of odd-parity perturbations 
in the form  
\be
S_{\rm odd}=\sum_{l,m} L \int dt dr\, 
{\cal L}_{\rm odd}\,, 
\label{oddact}
\ee
where $L=l(l+1)$, and 
\ba
{\cal L}_{\rm odd}&=&r^2 \sqrt{\frac{f}{h}}
\biggl[
C_1\left(\dot{W}_{lm}-Q'_{lm}+\frac{2}{r}Q_{lm}\right)^2+2\left(C_2\dot{\da}_{lm}
+C_3\da'_{lm}+C_4\da_{lm} \right)
\left(\dot{W}_{lm}-Q'_{lm}+\frac{2}{r}Q_{lm}\right)
+C_5\dot{\da}_{lm}^2
\notag\\
&&
+C_6\dot{\da}_{lm}\da'_{lm}
+C_7{\da}_{lm}'^2
+(L-2)\left( C_8W_{lm}^2+C_9 W_{lm}\da_{lm}
+\frac{A_0}{f}C_9W_{lm}Q_{lm}
+C_{10}Q_{lm}^2
+C_{11}Q_{lm}\da_{lm} \right)
\notag\\
&&
+(LC_{12}+C_{13})\da_{lm}^2 \biggr]\,,
\label{oddLag}
\ea
where a dot represents the derivative with respect to 
$t$, and 
\ba
C_1 &=& \frac{h}{2fr^2} \left[ G_4 -\frac{A_0^2-fh A_1^2}{f} 
G_{4,X}-\frac{(A_0^2-fh A_1^2)^2}{f^2}f_4 \right]\,,\nonumber \\
C_2 &=& -\frac{hA_1}{2f^2r^2} \left[ fG_{4,X}
+(A_0^2-fh A_1^2)f_4 \right]\,,\qquad
C_3 = \frac{hA_0}{2f^2r^2} \left[ fG_{4,X}
+(A_0^2-fh A_1^2)f_4 \right]\,,\nonumber \\
C_4 &=& \frac{1}{2fr^3} \biggl[ -hrA_0'+h (rA_0'-2A_0)G_{4,X} 
\nonumber \\
& &
+\frac{h}{f^2} (A_0 A_1^2 f^2 h' r+2A_0 A_1 A_1'f^2 h r
-A_0' A_1^2 f^2 h r+2A_0A_1^2 f^2 h+A_0^3f' r
-A_0^2A_0'fr-2A_0^3 f)f_4
\biggr]\,,\nonumber \\
C_5 &=& \frac{1}{2fr^2}\,,\qquad C_6=0\,,
\qquad C_7=-\frac{h}{2r^2}\,,\qquad 
C_8=-\frac{h}{2fr^4} \left[ f(G_4+h A_1^2 G_{4,X})
+hA_1^2 (A_0^2-fh A_1^2) f_4 \right]\,,\nonumber \\
C_9 &=& \frac{hA_1}{fr^4} \left[ fG_{4,X}
+(A_0^2-fh A_1^2)f_4  \right]\,,\qquad
C_{10}=\frac{1}{2f^3 r^4} \left[ f (fG_4-A_0^2 G_{4,X})
-A_0^2 (A_0^2-fh A_1^2) f_4 \right]\,,\nonumber \\
C_{11} &=& -\frac{A_0}{f^2 r^4} \left[ f G_{4,X}
+(A_0^2-fh A_1^2) f_4 \right]\,,\qquad
C_{12}=-\frac{1}{2r^4}\,.
\label{Cform}
\ea
Since the coefficient $C_{13}$ is not needed 
in the following discussion, we do not write 
its explicit expression here. 
The coefficient $C_{6}$ vanishes in quartic-order 
BGP theories, but this is not the case 
in the presence of other interactions \cite{KMTZ}.

We can derive conditions for the absence of ghosts 
and Laplacian instabilities by following the procedure 
given in Ref.~\cite{KMTZ}. There are two dynamically 
propagating modes: 
\be
\chi \equiv \dot{W}_{lm}-Q'_{lm}
+\frac{2}{r}Q_{lm}+\frac{C_2\dot{\da}_{lm}
+C_3\da'_{lm}+C_4\da_{lm}}{C_1}\,,\qquad 
\delta A_{lm}\,, 
\ee
for $l\geq2$. For the monopole mode ($l=0$), 
the Lagrangian (\ref{oddLag}) vanishes identically. 
For the dipole mode ($l=1$), the perturbation $\chi$ becomes 
non-dynamical and the vector-field perturbation $\delta A_{1m}$ is the only 
propagating DOF. As shown in Ref.~\cite{KMTZ}, the mode $\delta A_{1m}$ 
possesses the propagation speed same as that 
for $\delta A_{lm}(l\geq2)$ in GP theories.
Hence, the perturbation $\delta A_{lm}$ corresponds to 
the intrinsic vector mode, and consequently
the other mode $\chi$ is associated with 
the tensor perturbation arising from the gravity sector. 

Introducing $\chi$ as a Lagrange multiplier in the action 
and eliminating $W_{lm}$ and $Q_{lm}$ from $S_{\rm odd}$ 
by using their perturbation equations of motion, 
the second-order Lagrangian is expressed in the form 
\be
(L-2){\cal L}_{\rm odd}=r^2 \sqrt{\frac{f}{h}}\left( 
\dot{\vec{\mathcal{X}}}^{t}{\bm K}\dot{\vec{\mathcal{X}}}
+\dot{\vec{\mathcal{X}}}^{t}{\bm R}\vec{\mathcal{X}}'
+\vec{\mathcal{X}}'^{t}{\bm G}\vec{\mathcal{X}}'
+\vec{\mathcal{X}}^{t}{\bm M}\vec{\mathcal{X}}
\right)\,,
\label{LM2}
\ee
where $\vec{\mathcal{X}}^{t}=(\chi,\da_{lm})$, and 
${\bm {K,R,G,M}}$ are $2\times2$ matrices. 
In general, there are other contributions  
$\vec{\mathcal{X}}'^{t}{\bm S}\vec{\mathcal{X}}$ and 
$\dot{\vec{\mathcal{X}}}^{t}{\bm T}\vec{\mathcal{X}}$ 
to the Lagrangian ${\cal L}_{\rm odd}$ \cite{KMTZ}.
The diagonal components of matrices ${\bm S}$ and ${\bm T}$ 
can be absorbed into ${\bm M}$ after integration by parts. 
Moreover, the off-diagonal components 
of ${\bm S}$ and ${\bm T}$ vanish by 
using the coefficients given in Eq.~(\ref{Cform}). 
Hence the second-order Lagrangian in quartic-order 
BGP theories can be expressed 
in the form (\ref{LM2}) without 
the contributions $\vec{\mathcal{X}}'^{t}{\bm S}\vec{\mathcal{X}}$ 
and $\dot{\vec{\mathcal{X}}}^{t}{\bm T}\vec{\mathcal{X}}$.

The nonvanishing components of the kinetic matrix ${\bm K}$ 
are $K_{11}=q_1$ and $K_{22}=(L-2)q_2$, where
\be
q_1= \frac{4f^2C_1^2C_{10}}{A_0^2C_9^2-4f^2C_8C_{10}}\,,
\qquad
q_2=\frac{C_1C_5-C_2^2}{C_1}\,.
\label{q2def}
\ee
The sufficient conditions for the absence of ghosts 
correspond to $q_1>0$ and $q_2>0$. 

Let us first consider the radial propagation of odd-parity modes 
by assuming the solution 
of the form $\vec{\mathcal{X}}^{t} \propto e^{i(\omega t-kr)}$.
In the limit of large $\omega$ and $k$, the dispersion 
relation reduces to 
${\rm det}(\omega^2{\bm K}-\omega k {\bm R}+k^2 {\bm G})=0$.
The radial propagation speed $c_r$ 
in proper time is given by 
$c_r=\omega/(\sqrt{fh}\,k)$ \cite{KMTZ}. 
On using the fact that the nonvanishing components of 
${\bm R}$ and ${\bm G}$ are given by 
$R_{11}=A_0C_9K_{11}/(fC_{10})$, 
$R_{22}=-2C_2 C_3 (L-2)/C_1$, 
$G_{11}=C_8K_{11}/C_{10}$, and 
$G_{22}=(L-2)(C_1 C_7-C_3^2)/C_1$, we obtain
the two propagation speeds from the dispersion relation:
\be
c_{r1}=\frac{A_0C_9\pm
\sqrt{A_0^2C_9^2-4f^2C_8C_{10}}}{2f^{3/2}h^{1/2}C_{10}}\,, 
\qquad
c_{r2}=\frac{-2C_2C_3\pm
2\sqrt{C_1C_3^2C_5-C_1^2 C_7 q_2}}
{2f^{1/2}h^{1/2}C_1q_2}\,. 
\label{cr2}
\ee
The Laplacian instability along the radial direction 
can be avoided for $c_{r1}^2 \geq 0$ and 
$c_{r2}^2 \geq 0$.

For the modes $L \gg 1$, we substitute the solution 
$\vec{\mathcal{X}}^{t} \propto e^{i(\omega t-l \theta)}$ into 
Eq.~(\ref{LM2}) to derive propagation speeds 
along the angular direction. 
Then, the dispersion relation corresponds to 
${\rm det}(\omega^2{\bm K}+{\bm M})=0$. 
The leading-order diagonal components of the matrix ${\bm M}$ 
are $M_{11}=-LC_1$ and $M_{22}=L(L-2)D_1$, where
\be
D_1=C_{12}+\frac{fC_8C_{11}^2+C_9^2(fC_{10}-A_0C_{11})}
{4fC_1^2C_{10}}q_1\,.
\label{defD2}
\ee
The propagation speed squared along the angular direction 
in proper time is given by $c_{\Omega}^2=\omega^2r^2/(fl^2)$.
Taking the limit $L \to \infty$ in the dispersion relation, 
we obtain the two values:
\be
c_{\Omega 1}^2=\frac{C_1 r^2}{f q_1}\,,\qquad
c_{\Omega 2}^2=-\frac{D_1 r^2}{f q_2}\,.
\label{casq}
\ee
We require the two conditions $c_{\Omega 1}^2 \geq 0$ 
and $c_{\Omega 2}^2 \geq 0$ to avoid 
the Laplacian instability along the angular direction.
Since the matrices ${\bm K}$, ${\bm R}$, and ${\bm G}$
are diagonal and the matrix ${\bm M}$ also becomes 
diagonal in the limit $L\gg 1$,
the tensor mode $\chi$ and the intrinsic vector mode $\delta A_{lm}$
are orthogonal and decoupled in the high-frequency limit.

We recall that, under the condition (\ref{f4con}), 
the cosmological value of $c_t^2$ is equivalent to 1. 
We compute the quantities 
$q_1,q_2,c_{r1}^2,c_{r2}^2,c_{\Omega 1}^2,c_{\Omega 2}^2$ 
by imposing (\ref{f4con}). 
Since the condition (\ref{f4con}) translates to 
$fG_{4,X}+(A_0^2-fh A_1^2)f_4=0$, some of the 
coefficients in Eq.~(\ref{Cform}) reduce to 
\be
C_1=\frac{hG_4}{2fr^2}\,,\qquad 
C_2=C_3=C_9=C_{11}=0\,,\qquad 
C_8=-\frac{hG_4}{2r^4}\,,\qquad
C_{10}=\frac{G_4}{2fr^4}\,.
\label{Coeff}
\ee
In Ref.~\cite{Babi18_2}  it was 
argued that, if the Lagrangian contains cross 
terms of both the time and spatial derivatives 
($\dot{\vec{\mathcal{X}}}^{t}{\bm R}\vec{\mathcal{X}}'$ in our theory), 
the positivity of kinetic matrix ${\bm K}$ is not necessarily required for 
the Hamiltonian bounded from below. 
In other words, provided that the cross terms associated with the matrix ${\bm R}$ 
do not vanish, 
the two conditions $q_1>0$ and $q_2>0$ are sufficient but not necessary for the absence of ghosts. 
In our BGP theory the matrix components of ${\bm R}$ 
vanish identically by using Eq.~(\ref{Coeff}), so the 
sufficient conditions for the absence of ghosts 
translate to $q_1>0$ and $q_2>0$.
These quantities  yield
\be
q_1=-\frac{C_1^2}{C_8}=\frac{hG_4}{2f^2}\,,\qquad 
q_2=C_5=\frac{1}{2fr^2}\,.
\ee
Provided that $G_4>0$, the conditions $q_1>0$ and 
$q_2>0$ are trivially satisfied outside the horizon.

The squares of the radial propagation speeds 
in Eq.~(\ref{cr2}) are given by 
\be
c_{r1}^2=-\frac{C_8}{fh C_{10}}=1\,,\qquad 
c_{r2}^2=-\frac{C_7}{fh q_2}=1\,.
\ee
On using the fact that $D_1$ is equivalent to 
$C_{12}=-1/(2r^4)$, the propagation speed squares 
in the angular direction are
\be
c_{\Omega 1}^2=1\,,\qquad 
c_{\Omega 2}^2=1\,.
\ee

We have thus shown that, under the condition (\ref{f4con}), 
the propagation speeds for odd-parity perturbations 
on the static and spherically symmetric background
are all equivalent to 1.
The propagation speeds $c_{r1}$ and $c_{\Omega 1}$ 
can be identified with those arising from tensor perturbations. 
Then, under the condition (\ref{f4con}), 
the speed of gravitational waves propagating 
around BHs is the same as the cosmological value 
$c_t=1$. The other speeds $c_{r2}$ and $c_{\Omega 2}$ 
correspond to those arising from vector-field perturbations. 
For quartic-order BGP theories, 
the propagation speed squared of vector perturbations on the 
FLRW cosmological background is given by \cite{HKT16}
\be
c_v^2=1+\frac{2X(G_{4,X}+2Xf_4)^2}
{G_4-2XG_{4,X}-4X^2f_4}\,.
\ee
Under the condition (\ref{f4con}), it follows that 
$c_v^2=1$. This is consistent with the fact that 
both $c_{r2}^2$ and $c_{\Omega 2}^2$ 
are equivalent to 1 on the background (\ref{metric}). 
The coincidence of the propagation speed of the vector perturbation 
with that of the tensor perturbation and their coincidence 
with the speed of light
arises from the specific choice of our theory
(\ref{action}) with the condition (\ref{f4con}).
For instance, if the action (\ref{action})
contains nonlinear kinetic terms of the vector field
$G_2(X, F,Y)$ with $F=-F^{\mu\nu}F_{\mu\nu}/4$ and $Y=A^{\mu}A^{\nu}F_{\mu\rho}{F_{\nu}}^{\rho}$,
the propagation speed of vector perturbations 
generally differs from the speed of light,
while that of tensor perturbations remains the same.

It is also natural to expect that,  
if the propagation speed of a mode 
on the cosmological background 
coincides with the speed of light,
that on the BH background should also
coincide with the speed of light,
since the propagation speed of perturbations 
is locally fixed on scales much shorter than 
background curvature radii.
Thus, if the propagation speed of the vector mode 
on the cosmological background $c_v$
is equivalent to the speed of light,
those on the static and spherically symmetric background,
$c_{r2}^2$ and $c_{\Omega2}^2$,
also coincide with the speed of light.

For the dipole perturbation ($l=1$), 
only the vector perturbation $\delta A_{lm}$ 
propagates with the radial and angular speed 
squares $c_{r2}^2$ and $c_{\Omega 2}^2$, respectively. 
They are equivalent to 1 under the condition (\ref{f4con}). 

We note that the configuration of a linearly time-dependent 
scalar $\phi=qt+\psi(r)$ in quartic-order 
shift-symmetric Horndeski theories 
and its GLPV extension \cite{GLPV} can be recovered by taking 
the limits $\delta A_{lm} \to 0$, $A_0 \to q$, and 
$A_1 \to \psi'$, where $q$ is a constant and $\psi$ is 
a function of $r$. 
The fact that the condition (\ref{f4con}) 
is sufficient to guarantee the values $c_{r1}^2=c_{\Omega 1}^2=1$
in BGP theories means that the same result also holds 
in quartic-order shift-symmetric GLPV theories. 
Thus, we proved that the claim of Ref.~\cite{Babi18} 
is correct for odd-parity perturbations without putting 
any restriction on the models.

As we mentioned in Introduction, the charged stealth Schwarzschild solution arising from the specific quartic coupling $G_4(X)=M_{\rm pl}^2/2+X/4$ 
in GP theories is unstable against odd-parity perturbations in the vicinity of the event horizon \cite{KMTZ}.
We note that, by the ``charged stealth Schwarzschild''
solution \cite{Chagoya}, we distinguish it from 
the ``stealth Schwarzschild'' solution 
obtained in Ref.~\cite{Babi14} and 
its straightforward extension to the GP theory with $G_4=M_{\rm pl}^2/2+\beta X$
and $F_{\mu\nu}=0$ \cite{Minami16},
where $\beta$ is an arbitrary dimensionless coupling constant.
One may wonder if this instability can be 
alleviated according to the discussion of no-ghost criterion claimed 
in Ref.~\cite{Babi18_2}. 
As we will show in Appendix~\ref{appendix2}, 
this is not the case since the origin of this instability is not the appearance of 
ghosts but the propagation speed squared being negative. 
Thus, the conclusion of Ref.~\cite{KMTZ}
was rather obtained from the same criterion as 
the hyperbolicity condition employed in Ref.~\cite{Babi18_2}.
This charged stealth Schwarzschild solution
has a nonzero electric field
and hence there is no counterpart solution in scalar-tensor theories 
obtained by the replacement of $A_\mu$ with $\partial_\mu\phi$. 
Thus, our argument here is peculiar for vector-tensor theories.

Finally, one may concern that the instability of the stealth Schwarzschild solution 
stemming from the model $G_4(X)=M_{\rm pl}^2/2+X/4$ 
in GP theories \cite{KMTZ} would contradict with the stability of 
our model in BGP theories described by the action (\ref{action}), 
as these two theories may be related to each other
via a disformal transformation. 
As we show in Appendix~\ref{appendix3}, however, 
the disformal transformation cannot exactly map the former into the latter. 
After the transformation, there are new interactions
of the forms (\ref{al67}) \cite{KNY16}. 
Hence the quadratic GP theory after the disformal transformation  is not 
physically equivalent to our BGP theory given by the action (\ref{action}).

\section{Hairy BH solutions}
\label{BHsec}

In this section, we derive hairy BH solutions in quartic-order 
BGP theories.
The background equations of motion (\ref{be1})-(\ref{be4}) 
can be expressed in the form
\be
Z {\bm x}={\bm y}\,,
\label{Zeq}
\ee
where ${\bm x}={}^t (f', h', A_0'', A_1')$, 
$Z$ and ${\bm y}$ are $4 \times 4$ and $1 \times 4$ 
matrices, respectively, which contain the dependence 
of $f, h, A_0', A_0, A_1$. 
The components $Z_{11},Z_{13},Z_{22},Z_{23},Z_{43},Z_{44}$ 
of the matrix $Z$ vanish, so the determinant of $Z$ 
reduces to ${\rm det}\,Z=Z_{33}(Z_{12}Z_{24}Z_{41}
+Z_{21}Z_{42}Z_{14})$.
On using the relations 
$Z_{21}=-(h/f)Z_{12}$, 
$Z_{42}=-Z_{24}/(2h)$, and $Z_{41}=-Z_{14}/(2f)$,  
it follows that 
\be
{\rm det}\,Z=0\,.
\ee
Hence we cannot solve Eq.~(\ref{Zeq}) for ${\bm x}$ 
to derive closed-form differential equations. 
This property generally holds in quartic-order 
BGP theories on the static and spherically symmetric background 
without imposing the condition (\ref{f4con}). 

We note that the determinant also vanishes for the dynamics of 
anisotropic cosmology in quartic-order BGP 
theories \cite{ani}. 
Then, the property of vanishing determinant arises for 
vector-tensor theories with the equations of motion 
higher than second order under the metric ansatz with
maximally-symmetric two-dimensional space. 
It is an open question whether such behavior 
generally occurs in the spacetime with the two-dimensional 
maximally-symmetric space for other gravitational 
theories beyond second order (e.g., GLPV theories), 
which we would like to address in a future publication.

The fact that the background equations of motion are not closed means 
that we need additional conditions to close the system.
{}From Eq.~(\ref{be4}), there are in general two branches: 
(a) $A_1=0$, or (b) $A_1 \neq 0$. 

For the branch (a), Eq.~(\ref{be4}) is redundant, so the differential 
equations (\ref{Zeq}) reduce to the system of the $3 \times 3$ 
matrix $Z$ with ${\bm x}={}^t (f', h', A_0'')$. 
In this case, the determinant of $Z$ is given by 
\be
{\rm det}\,Z=-\frac{4h^2}{r^2f^4} 
\left( A_0^2 G_{4,X}-fG_4 \right)^2\,,
\ee
which does not generally vanish. 
Then, we can solve Eq.~(\ref{Zeq}) for the variables 
$f, h, A_0$. In Sec.~\ref{nusec}, we will obtain 
numerical BH solutions for the branch $A_1=0$ by 
considering quartic-order power-law couplings.

For the branch (b), we need to impose at least one condition 
to close the system (\ref{Zeq}). 
In Refs.~\cite{HKMT}, the authors found exact BH solutions 
in GP theories by imposing the two conditions $f=h$ and 
$X={\rm constant}$.
In Sec.~\ref{exactsec}, we will find exact BH solutions 
in quartic-order BGP theories by imposing the same conditions.

\subsection{Numerical solutions for the branch $A_1=0$}
\label{nusec}

In this subsection, we will focus on the branch 
\be
A_1=0\,,
\ee
and numerically obtain hairy BH solutions  
for power-law couplings  
\be
G_4(X)=\frac{M_{\rm pl}^2}{2}
+\beta_4 M_{\rm pl}^2 \left(\frac{X}{M_{\rm pl}^2}
\right)^n,
\label{G4X}
\ee
where $n\geq 1$ is an integer and $\beta_4$ is a constant.
We also impose the condition (\ref{f4con}), under which 
the function $f_4$ is given by 
\be
f_4(X)=-\frac{n\beta_4}{2M_{\rm pl}^2}
\left(\frac{X}{M_{\rm pl}^2}\right)^{n-2}\,.
\label{f4X}
\ee

Around the event horizon characterized by the distance $r_h$, 
we iteratively derive the solutions to Eqs.~\eqref{be1}-\eqref{be3} 
by using the expansions:
\be
f=\sum_{i=1}^\infty f_i (r-r_h)^i\,,
\qquad
h=\sum_{i=1}^\infty h_i (r-r_h)^i\,,
\qquad
A_0=a_0+\sum_{i=1}^\infty a_i (r-r_h)^i\,,
\label{fhA0}
\ee
where $f_i, h_i, a_0$ are constants. 
The coupling $\beta_4$ works as corrections to the 
metric components of the Reissner-Nordstr\"{o}m (RN) 
solution: $f_{\rm RN}=h_{\rm RN}=(1-r_h/r)(1-\mu r_h/r)$, 
where $\mu$ is a constant in the range $0<\mu<1$.
Substituting Eq.~(\ref{fhA0}) into Eqs.~\eqref{be1}-\eqref{be3} 
for the branch $A_1=0$, the leading-order coefficients are 
given by  
\be
f_1=h_1=\frac{1-\mu}{r_h}\,,
\qquad
a_0=0,
\qquad
a_1=\frac{\sqrt{2\mu}\,M_{\rm pl}}{r_h}\,,
\label{f1h1}
\ee
where we have assumed $f_1=h_1$. 
The result (\ref{f1h1}) holds irrespective of the values 
of $n$, but the next-to-leading order coefficients depend on 
the power $n$.

For $n=1$, the nontrivial $\beta_4$ dependence appears 
at the order of ${\cal O}((r-r_h)^2)$, as 
\be
f_2=-
\frac{1-(3-4\beta_4)\mu + (2-5\beta_4)\mu^2}
       {(1-\mu)r_h^2},
\qquad
h_2
=-\frac{1-3\mu+(2+3\beta_4)\mu^2}
      {(1-\mu)r_h^2},
\qquad
a_2
=
-\frac{\sqrt{2\mu}\,M_{\rm pl} 
\left[(1-\mu)^2-\beta_4 \mu^2\right]}
{(1-\mu)^2r_h^2}.
\ee

For $n=2$, the coupling $\beta_4$ appears at the 
order of ${\cal O}((r-r_h)^3)$, as
\ba
f_2
&=& 
h_2
=-\frac{1-2\mu}{r_h^2},
\qquad
a_2
=-\frac{\sqrt{2 \mu}\,M_{\rm pl}}{r_h^2}\,,
\nonumber \\
f_3
&=&
\frac{3-15\mu +3 (7-4\beta_4)\mu^2+ (14\beta_4-9)\mu^3}
      {3(1-\mu)^2r_h^3},
\qquad
h_3
=
\frac{3-15\mu +21\mu^2-(9+10\beta_4)\mu^3}
      {3(1-\mu)^2r_h^3}\,,
      \nonumber \\
a_3
&=&
\frac{\sqrt{2\mu}M_{\rm pl}
        [3-9\mu +9\mu^2 +(2\beta_4-3)\mu^3]}     
      {3(1-\mu)^3 r_h^3}.
\ea

For $n>2$, the nontrivial $\beta_4$ dependence 
around the horizon appears at the order of 
${\cal O}((r-r_h)^{n+1})$. 
Thus, the regularity of $f,h,A_0$ at the horizon is ensured 
for general $n~(\geq 1)$.

At large distances ($r\gg r_h$), the iterative 
solutions for general $n$ are given by 
\ba
f&=& 1
-\frac{2^{n+1}M (M_{\rm pl}/P)^{2n} 
+4M \beta_4 +8n Q\beta_4/P}
{\left[2^n (M_{\rm pl}/P)^{2n}
-2(2n-1)\beta_4\right]r}
+{\cal O} (r^{-2}),
\\
h&=&1
-\frac{2M}{r}
+\frac{2^{n-1}Q^2}
{M_{\rm pl}^2 \left[2^n-2(2n-1) 
(P/M_{\rm pl})^{2n}\beta_4 \right]r^2}
+{\cal O} (r^{-3}),
\\
A_0
&=&
P+\frac{Q}{r}
-\frac{nQ (2MP+Q)\beta_4}
{P\left[2^n (M_{\rm pl}/P)^{2n}
-2(2n-1)\beta_4\right]r^2}
+{\cal O}(r^{-3}).
\label{fasy}
\ea
The coupling $\beta_4$ works as corrections to the 
RN solution with $A_0=P+Q/r$.

\begin{figure}[h]
\begin{center}
\includegraphics[height=3.3in,width=3.3in]{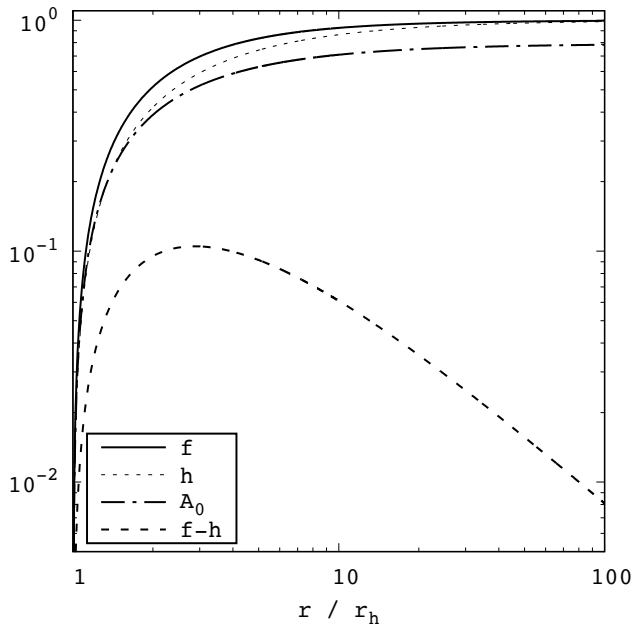}
\end{center}
\caption{\label{fig1}
Numerical solutions to $f,h,A_0,f-h$  
for the couplings (\ref{G4X}) and (\ref{f4X}) 
with $n=2$, $\beta_4=0.49$ and $\mu=0.3$.
The boundary conditions are chosen to be consistent 
with the expansion (\ref{fhA0}) 
at the distance $r=1.001r_h$. 
The temporal vector component $A_0$ is normalized 
by $M_{\rm pl}$. 
The solutions are regular throughout the horizon exterior.}
\end{figure}

In Fig.~\ref{fig1}, we plot the numerically integrated solutions 
of $f,h,A_0,f-h$ for $n=2$, $\beta_4=0.49$, and $\mu=0.3$. 
We employ the iterative solutions (\ref{fhA0}) up to third 
order as boundary conditions in the vicinity of the horizon 
and solve Eqs.~(\ref{be1})-(\ref{be3}) for the branch $A_1=0$.
The two asymptotic solutions in the regimes $r \simeq r_h$ and 
$r \gg r_h$ smoothly connect to each other without any 
discontinuity. As estimated above, the temporal vector component 
$A_0$ is close to 0 around the horizon and then it increases 
toward the asymptotic value $P$ as $r \to \infty$.

We also numerically confirmed that the curvature invariants such as $R$, $R_{\mu\nu}R^{\mu\nu}$, and $R_{\mu\nu\alpha\beta}R^{\mu\nu\alpha\beta}$ (where 
$R_{\mu \nu}$ is the Ricci tensor and 
$R_{\mu\nu\alpha\beta}$ is the Riemann tensor) 
are regular at/outside the horizon and hence there is no curvature singularity. 
Using the iterative solutions (\ref{fhA0}) for $n=2$
and picking up the dominant contributions around 
the BH event horizon, these quantities reduce to 
$R\to [20\mu^2\beta_4/\{(1-\mu)r_h^3\}](r-r_h)$,
$R_{\mu\nu}R^{\mu\nu}\to 4\mu^2/r_h^4$,
and $R_{\mu\nu\alpha\beta}R^{\mu\nu\alpha\beta}\to
4(5\mu^2-6\mu+3)/r_h^4$,
while they converge to 0 
at spatial infinity. 
We note that for general $n~(\geq 1)$, $R\sim (r-r_h)^{n-1}$,
while $R_{\mu\nu}R^{\mu\nu}$ and $R_{\mu\nu\alpha\beta}R^{\mu\nu\alpha\beta}$
approach constant as $r\to r_h$.

In our numerical simulation, we have shifted 
the value of $f$ to 1 at the distance $r=10^7r_h$ by using the 
freedom of time rescaling. 
In Fig.~\ref{fig1}, we observe that 
the difference between $f$ and $h$ induced by the coupling 
$\beta_4$ is most significant in the vicinity of the horizon 
($f-h \simeq 0.1$ around $r \simeq 3r_h$). 
This difference may be potentially probed in future high-precision 
GW measurements in nonlinear regimes of gravity. 

We have thus shown the existence of hairy BH solutions 
regular throughout the horizon exterior for $n=2$.
Numerically, we have also confirmed that the two asymptotic 
solutions (\ref{fhA0}) and (\ref{fasy}) are smoothly joined 
each other for general powers of $n~(\geq 1)$.
Since there are two independent parameters 
$r_h$ and $\mu$ for the near-horizon solutions, 
the charge $P$ generally 
depends on $M$ and $Q$. Hence the Proca hair $P$ is of 
the secondary type.

\subsection{Exact BH solutions}
\label{exactsec}

The exact BH solution found for the specific coupling 
$G_4(X)=M_{\rm pl}^2/2+X/4$ in GP theories \cite{Chagoya} 
satisfies the two relations
\be
f=h,
\qquad
X=X_c,
\label{twocon}
\ee
where $X_c$ is a constant. In the following, we will search 
for exact BH solutions in quartic-order BGP theories by imposing 
the two conditions (\ref{twocon}). 
The second condition gives the relation 
$A_1^2=(A_0^2-2fX_c)/(fh)$ between $A_1$ and $A_0$ \footnote{ 
Here we note that the longitudinal mode $A_1$ diverges at the horizon 
where $f=0$ as long as $A_1\neq0$. This behavior is simply comes from 
the choice of coordinate. In fact, one can show that the product 
$A_{\mu}dx^{\mu}$ is regular at the future and past event horizons 
by introducing the advanced and retarded null coordinates with 
the tortoise coordinate; see Ref.~\cite{Minami16} and also 
Refs.~\cite{HKMT,HKMT2}.}. 

{}From Eq.~\eqref{be4}, it follows that  
\be
\left[
A_0^2+2r A_0 A_0' 
-X_c (1+f+rf')
\right]G_{4,X}(X_c)\,A_1=0\,,
\label{be4_red}
\ee
so there are three branches satisfying 
(i) $A_0^2+2r A_0 A_0' -X_c (1+f+rf')=0$, 
(ii) $G_{4,X}(X_c)=0$, and 
(iii) $A_1=0$.

\subsubsection{Branch {\rm (i)}}

For this branch, the derivative $f'$ is given by 
\be
f'=\frac{A_0^2+2r A_0 A_0'-X_c (1+f)}{X_c r}\,.
\label{dfeq}
\ee
Substituting this relation into Eq.~(\ref{be3}), 
we obtain
\be
A_0''+\frac{2}{r}A_0=0\,,
\label{a0_eq}
\ee
whose integrated solution is 
\be
A_0=P+\frac{Q}{r}\,,
\label{A0so}
\ee
where $P$ and $Q$ are constants.
Substituting Eqs.~(\ref{dfeq}) and (\ref{A0so}) 
into Eq.~(\ref{be1}), it follows that 
\be
4 \left( P^2-2X_c \right) G_4(X_c)r^2
+\left[X_c-4G_4(X_c)\right]Q^2=0\,,
\label{Preq}
\ee
under which Eq.~(\ref{be2}) is also satisfied.
To ensure the equality of Eq.~(\ref{Preq}) for 
arbitrary $r$, we require the two conditions
\be
X_c=\frac{P^2}{2}\,,\qquad
\left[ X_c-4G_4(X_c)\right] Q^2 =0\,.
\ee
The second condition is satisfied for either 
(A) $G_4(X_c)=X_c/4=P^2/8$, or (B) $Q=0$.

In the case (A), Eq.~(\ref{dfeq}) reduces to 
$f'=[P^2(1-f)r^2-2Q^2]/(P^2 r^3)$, which is 
integrated to give 
\be
f=h=1-\frac{2M}{r}+\frac{2Q^2}{P^2r^2}\,,\qquad 
A_1=\pm \frac{\sqrt{2P(MP+Q)r-Q^2}}{rf}\,,
\label{exact}
\ee
where $M$ is an integration constant. 
If we identify the constant $P$ as $2M_{\rm pl}$, 
the metric components in Eq.~(\ref{exact}) reduce 
to the RN solution with $G_4(X_c)=M_{\rm pl}^2/2$.
The difference from the RN solution in GR is that 
there is a nonvanishing longitudinal mode $A_1$. 
We require that $2P(MP+Q)r>Q^2$ for the existence 
of the exact solution (\ref{exact}). 
At spatial infinity, the longitudinal mode decreases 
as $A_1 \propto 1/\sqrt{r}$. 
The solution (\ref{exact}) exists for the couplings
\be
G_4(X)=\frac{P^2}{8}
+\sum_{n=1}^{\infty} b_n 
\left(X-\frac{P^2}{2}\right)^n\,,\qquad
f_4(X)
=-\frac{1}{2X}\sum_{n=1}^{\infty} n b_n 
\left(X-\frac{P^2}{2}\right)^{n-1}\,,
\label{branch1}
\ee
where $b_n$ are arbitrary constants.

The case (B) corresponds to the special case of (A), 
i.e., $Q=0$ in Eq.~(\ref{exact}).
Namely, this is the stealth Schwarzschild solution 
$f=h=1-2M/r$ with $A_0=P$ and $A_1=P\sqrt{2M/r}/f$.
This solution exists for arbitrary regular functions 
$G_4(X)$.

\subsubsection{Branch {\rm (ii)}}

We proceed to the second branch characterized by $G_{4,X}(X_c)=0$.
In this case, Eq.~(\ref{be3}) reduces to (\ref{a0_eq}), so 
the solution to $A_0$ is given by Eq.~(\ref{A0so}). 
{}From Eq.~(\ref{be1}), we obtain
\be
Q^2+4r^2 G_4(X_c) \left( rf'+f-1 \right)=0\,,
\ee
under which Eq.~(\ref{be2}) is also satisfied.
This gives the following integrated solution
\be
f=h=1-\frac{2M}{r}+\frac{Q^2}{4G_4(X_c)r^2}\,,
\qquad 
A_1=\pm \frac{1}{rf}
\sqrt{\frac{[(2P^2-4X_c)r^2+(4PQ+8MX_c)r+2Q^2]G_4(X_c)
-Q^2 X_c}{2G_4(X_c)}}\,,
\label{exact2}
\ee
with $A_0=P+Q/r$. Provided that $X_c \neq P^2/2$, the 
longitudinal mode $A_1$ approaches a constant 
for $r \to \infty$. This behavior is different from 
the branch (i) in which $A_1$ decreases toward 0 
due to the condition $X_c=P^2/2$.
The exact solution (\ref{exact2}) can be realized for 
the couplings
\be
G_4(X)=G_4(X_c)+\sum_{n=2}^{\infty} 
b_n (X-X_c)^n\,,\qquad
f_4(X)=-\frac{1}{2X}
\sum_{n=2}^{\infty} n b_n (X-X_c)^{n-1}\,.
\label{branch2}
\ee
If we choose $G_4(X_c)=M_{\rm pl}^2/2$, the 
metric components $f$ and $h$ in Eq.~(\ref{exact2}) are 
the same as those of the RN solution.

\subsubsection{Branch {\rm (iii)}}

Let us finally discuss exact solutions for the branch (iii) 
satisfying $A_1=0$. In this case, the two conditions 
(\ref{twocon}) give $A_0=\sqrt{2fX_c}$, where we have 
chosen the branch $A_0>0$.
Multiplying Eqs.~(\ref{be1}) and (\ref{be2}) by 
$G_4(X_c)$ and $G_4(X_c)-2X_c G_{4,X}(X_c)$, respectively, 
and taking their sums, it follows that 
\be
X_c^2 f'^2 G_{4,X} (X_c)=0\,.
\ee
Since we are considering the case $X_c \neq 0$, 
we obtain
\be
G_{4,X}(X_c)=0.
\ee
Then, Eq.~(\ref{be3}) reduces to 
\be
rf'^2 -2f \left(2f'+rf'' \right)=0\,,
\ee
which is integrated to give
\be
f= \frac{C_1}{r^2} \left( \frac{r}{M}-1 
\right)^2\,,
\ee
where $C_1$ and $M$ are constants.
{}From Eq.~\eqref{be1}, we obtain 
\be
2G_4(X_c) \left( C_1-M^2 \right)r^2
+\left[ X_c-2G_4(X_c) \right]C_1 M^2=0\,,
\ee
which also follows from Eq.~\eqref{be2}.
This relation is satisfied for
\be
C_1=M^2\,,\qquad
G_4(X_c)=\frac{X_c}{2}\,.
\ee
Then, the resulting solution is 
\be
f=h=\left(1-\frac{M}{r}\right)^2\,,\qquad
A_0=\sqrt{2X_c} \left( 1-\frac{M}{r} \right)\,,
\qquad
A_1=0\,,
\label{so14}
\ee
which corresponds to the extremal RN solution.
The above exact solution can be realized by the 
couplings
\be
G_4(X)=
\frac{X_c}{2}+\sum_{n=2}^{\infty} b_n 
(X-X_c)^n\,,\qquad
f_4(X)=-\frac{1}{2X} 
\sum_{n=2}^{\infty} n b_n \left(X-X_c \right)^{n-1}\,.
\label{model3}
\ee
The solution (\ref{so14}) is the special case of 
Eq.~(\ref{exact2}) with the correspondence
\be
G_4(X_c)=\frac{X_c}{2}\,,\qquad
P=\sqrt{2X_c}\,,\qquad 
Q=-\sqrt{2X_c}M\,,
\ee
under which $A_1$ vanishes.

\section{Conclusions}
\label{consec}

The recent event GW170817 showed that the GW speed
$c_t$ traveling over the cosmological distance is 
very close to 1.
This fact put strong constraints on models of 
cosmic acceleration in the framework of modified gravity theories.
In GP theories with second-order equations of motion, the 
quartic- and quintic-order interactions are not allowed, 
unless their coupling constants are very small. 
In the healthy extension of GP theories (dubbed BGP theories), 
the additional quartic-order interaction (\ref{L4BGP}) 
gives rise to a model in which the cosmological value of $c_t$ 
is equivalent to 1 under the condition (\ref{f4con}).

The remaining question is whether the condition (\ref{f4con}) 
is sufficient to ensure that the speed of GWs around massive bodies 
like BHs is equal to 1 as well. To address this point, we considered 
metric and vector-field perturbations in the odd-parity 
sector on the static and spherically symmetric background 
in quartic-order BGP theories.
We explicitly showed that, under the condition (\ref{f4con}), 
the propagation speeds $c_{r1}$ and $c_{\Omega 1}$ along the 
radial and angular directions in the gravity sector are 
both equivalent to 1. Under the same condition, we also found 
that the speeds of vector-field perturbations in 
the radial and angular directions reduce to 1. 
The no-ghost conditions are trivially satisfied for $G_4>0$. 
Our result about the GW speed around BHs 
is also valid in quartic-order 
shift-symmetric Horndeski and GLPV theories 
with the time-dependent scalar field $\phi=qt+\psi(r)$,
where $r$ is the radial coordinate, by taking 
the limits $\delta A_{lm} \to 0$, $A_0 \to q$, and 
$A_1 \to \psi'$. Hence we proved the claim of Ref.~\cite{Babi18} 
for odd-parity perturbations without restricting models.

We also searched for hairy BH solutions in quartic-order BGP 
theories by imposing the condition (\ref{f4con}). 
In general, the additional interaction beyond the domain of 
GP theories leads to a vanishing determinant for the 
equations of motion on the static and spherically symmetric 
background. This property does not hold under 
additional conditions, say, by choosing a branch 
with the vanishing longitudinal component ($A_1=0$) or by 
imposing the condition $f=h$.

For the branch $A_1=0$, we analytically derived iterative 
solutions around the horizon and at spatial infinity for 
the quartic-order power-law model (\ref{G4X}) with the BGP 
interaction (\ref{f4X}). Numerically, we also confirmed that 
the solutions in two asymptotic regimes connect to each other 
without any discontinuity outside the horizon.
The coupling $\beta_4$ works as corrections to the 
RN metric. As we see in Fig.~\ref{fig1}, the difference 
between two metric components $f$ and $h$ induced by 
$\beta_4$ is most significant in the vicinity of the horizon.

Imposing the two conditions $f=h$ and $X=X_c={\rm constant}$, 
we also obtained three branches of exact solutions in 
quartic-order BGP theories satisfying the condition 
(\ref{f4con}). The branch (i) corresponds to the RN-type 
solution (\ref{exact}) present for the model (\ref{branch1}), 
in which case the longitudinal mode has the dependence 
$A_1 \propto 1/\sqrt{r}$ at spatial infinity. 
The branch (ii) arises for the model (\ref{branch2}) 
with the RN-type metric given in Eq.~(\ref{exact2}), 
but $A_1$ approaches a constant for $r \to \infty$. 
The branch (iii), which exists for the model (\ref{model3}), 
corresponds to $A_1=0$ with the extremal RN metric 
given in Eq.~(\ref{so14}).

In GP theories with the quartic power-law coupling 
(\ref{G4X}), the branch $A_1 \neq 0$ is unstable 
against odd-parity perturbations \cite{KMTZ}. 
Moreover, the branch with $A_1=0$ gives rise to 
the speed of GWs approaching a constant different 
from 1 at spatial infinity, so this behavior can be 
odd with the observational bound of $c_t$. 
In contrast, all the numerical and exact BH solutions 
derived in this paper satisfy $c_t=1$ even in the 
vicinity of BHs, so they are not prone to the 
instability problem against odd-parity perturbations. 
Thus, the extension from GP theories to BGP theories 
allows the possibility for realizing hairy BH solutions 
in which the behavior of tensor perturbations is 
similar to that in GR.

In this paper, we focused on perturbations in the odd-parity sector, but it is necessary to study the behavior of even-parity perturbations in order to ensure the stability of BHs in the 
model with $c_t=1$.
In particular, the existence of scalar perturbations in the even-parity sector 
may give rise to additional constraints on the model parameters. 
The numerical solutions with $A_1=0$ and exact solutions with $Q\neq0$ 
presented in Secs.~\ref{nusec} and \ref{exactsec} do not exist as 
the counterparts of shift-symmetric Horndeski theories, so it is of interest 
to investigate the stabilities of them against even-parity perturbations.
It is also interesting to place observational constraints 
on dark energy models in quartic-order BGP theories 
satisfying the condition (\ref{f4con}). 
These issues are left for future works.

\appendix

\section{Coefficients in the background
equations of motion}
\label{appendix1}

The coefficients appearing in Eqs.~(\ref{be1})-(\ref{be4}) 
are given, respectively, by 
\ba
\hspace{-0.5cm}
c_1 &=& -2G_4+2 \left( \frac{A_0^2}{f}-2hA_1^2 \right) G_{4,X}
-\frac{2hA_1^2}{f} \left( A_0^2-fh A_1^2 \right) G_{4,XX} 
\nonumber \\
\hspace{-0.5cm}
& &
-\frac{2hA_1^2}{f} \left( 7A_0^2-5fh A_1^2 \right) f_4
-\frac{2h A_1^2}{f^2} \left( A_0^2-fh A_1^2 \right)^2 f_{4,X}\,,
\nonumber \\
\hspace{-0.5cm}
c_2 &=& -\frac{h}{2f}A_0'^2\,, \nonumber \\
\hspace{-0.5cm}
c_3 &=& -4h^2 A_1 A_1' G_{4,X}
-\frac{4h^2A_1}{f} \left( A_0^2 A_1'+A_0A_0'A_1
-fh A_1^2A_1' \right)G_{4,XX} \nonumber \\
\hspace{-0.5cm}
& &
-\frac{4h^2A_1}{f} \left( 5A_0^2 A_1'+3A_0A_0'A_1
-4fh A_1^2A_1' \right)f_4
-\frac{4h^2A_1}{f^2} \left( A_0^2-fh A_1^2 \right) 
\left( A_0^2 A_1'+A_0 A_0'A_1-fh A_1^2 A_1' \right)f_{4,X},
\nonumber\\
\hspace{-0.5cm}
c_4 &=& 2(1-h)G_4+\frac{2}{f} \left( h A_0^2-A_0^2 
-fh^2 A_1^2 \right)G_{4,X}
-\frac{2h^2A_0^2A_1^2}{f}G_{4,XX}
\nonumber \\
\hspace{-0.5cm}
& &
-\frac{2h^2A_1^2}{f} \left( 5A_0^2-fh A_1^2 \right)f_4
-\frac{2h^2A_0^2A_1^2}{f^2} 
\left( A_0^2-fh A_1^2 \right)f_{4,X}\,,\nonumber \\
\hspace{-0.5cm}
c_5 &=& \frac{h}{2f}A_0'^2\,,\nonumber \\
\hspace{-0.5cm}
c_6 &=& \frac{4h A_0A_0'}{f} \left( G_{4,X} 
-h A_1^2 G_{4,XX} \right)
+\frac{4h^2A_0A_1}{f} \left( A_0A_1'-5A_0'A_1 \right) f_4
-\frac{4h^2 A_0 A_0'A_1^2}{f^2} 
\left( A_0^2 -fh A_1^2 \right) f_{4,X}\,,\nonumber \\
\hspace{-0.5cm}
c_7 &=& 2\left( h-1 \right)G_4+2h (2h-1)A_1^2 G_{4,X} 
-2h^3 A_1^4 G_{4,XX}
+\frac{2h^2A_1^2}{f} \left( A_0^2-5fh A_1^2 \right)f_{4}
-\frac{2h^3A_1^4}{f} \left( A_0^2-fh A_1^2 \right)f_{4,X},
\nonumber\\
\hspace{-0.5cm}
d_1 &=& \frac{h}{2f^2}A_0'\,,\nonumber\\
\hspace{-0.5cm}
d_2 &=& \frac{2h^2A_0A_1^2}{f^2}G_{4,XX}
+\frac{6h^2A_0A_1^2}{f^2}f_4
+\frac{2h^2A_0A_1^2}{f^3} \left( A_0^2-fh A_1^2 
\right) f_{4,X}\,,\nonumber\\
\hspace{-0.5cm}
d_3 &=& -\frac{A_0'}{2f}\,,\nonumber\\
\hspace{-0.4cm}
d_4 &=& -\frac{2A_0}{f}G_{4,X}+\frac{2hA_0A_1^2}{f} 
G_{4,XX}+\frac{10h A_0A_1^2}{f}f_4
+\frac{2hA_0A_1^2}{f^2}\left( A_0^2-fh A_1^2 
\right) f_{4,X}\,,\nonumber\\
\hspace{-0.5cm}
d_5 &=& -\frac{h}{f}A_0''\,,\nonumber\\
\hspace{-0.5cm}
d_6 &=& -\frac{2hA_0'}{f}
+\frac{4h^2A_0A_1A_1'}{f}G_{4,XX}
+\frac{16h^2A_0A_1A_1'}{f}f_4 
+\frac{4h^2A_0A_1A_1'}{f^2}
\left( A_0^2-fh A_1^2 \right) f_{4,X}\,,\nonumber\\
\hspace{-0.5cm}
d_7 &=& -\frac{2(h-1)A_0}{f}G_{4,X}
+\frac{2h^2A_0 A_1^2}{f}G_{4,XX}
+\frac{8h^2A_0A_1^2}{f}f_4
+\frac{2h^2A_0A_1^2}{f^2}
\left( A_0^2-fh A_1^2 \right) f_{4,X}\,,\nonumber\\
\hspace{-0.5cm}
d_8 &=& \frac{2h^2A_1}{f}G_{4,X}
+\frac{2h^2A_1}{f^2} \left( A_0^2-fh A_1^2 
\right)G_{4,XX}
+\frac{2h^2A_1}{f^2} \left( 5A_0^2-4fh A_1^2 
\right)f_4
+\frac{2h^2A_1}{f^3} \left( A_0^2-fh A_1^2 
\right)^2f_{4,X}\,,\nonumber\\
\hspace{-0.5cm}
d_9 &=& -\frac{4h^2A_0A_0'A_1}{f}G_{4,XX}
-\frac{2hA_0A_1}{f} \left( 8h A_0'+h'A_0 \right)f_4
-\frac{4h^2A_0A_0'A_1}{f^2}\left( A_0^2-fh A_1^2 
\right) f_{4,X}\,,\nonumber\\
\hspace{-0.5cm}
d_{10} &=& 2h (h-1)A_1G_{4,X}-2h^3A_1^3 G_{4,XX}
-8h^3A_1^3 f_4
-\frac{2h^3A_1^3}{f} \left( A_0^2-fh A_1^2 
\right) f_{4,X}\,.\nonumber
\ea
%

\section{Instability of the charged stealth Schwarzschild solution 
in a specific GP theory}
\label{appendix2}

Let us briefly revisit the instability of the charged stealth Schwarzschild 
solution stemming from the specific quartic-order coupling in GP theories 
found in Ref.~\cite{KMTZ}. 
This solution arises for the couplings
\be
G_4=\frac{\Mpl^2}{2}+\frac14 X\,,\qquad 
f_4=0\,. 
\label{model}
\ee
In this model, there exists the following charged 
stealth Schwarzschild solution \cite{Chagoya}:
\be
f=h=1-\frac{2M}{r}\,,\qquad 
A_0=P+\frac{Q}{r}\,,\qquad 
A_1=\epsilon \frac{\sqrt{2P(MP+Q)r+Q^2}}{r-2M}\,, 
\label{stealthsol}
\ee
with $X=P^2/2$, where $M$, $P$, and $Q$ are integration constants. 
We substitute Eqs.~(\ref{model}) and (\ref{stealthsol}) into one of 
the propagation speed squares along the angular direction $c_{\Omega 1}^2$ 
given in Eq.~(\ref{casq}). Expanding it around the BH horizon at $r=2M$ 
and picking up the leading-order contribution, we obtain 
\be
c_{\Omega 1}^2=-\frac{M(4\Mpl^2+P^2)}{(2MP+Q)^2}(r-2M)
+{\cal O}((r-2M)^2)\,.
\ee
Provided that $M>0$, we have $c_{\Omega 1}^2<0$
outside the horizon.
Thus, the instability of BHs \eqref{stealthsol} arises due to the negative 
sound speed squared.

\section{Disformal transformation of a specific GP theory}
\label{appendix3}

In Appendix A.3 of Ref.~\cite{KNY16}, the disformal transformation of 
quartic-order GP theories is presented. 
Let us consider the theory given by the action  
\be
S=\int d^{4}x \sqrt{-\bar{g}} 
\left[ -\frac{1}{4}\bar{F}_{\mu \nu}\bar{F}^{\mu \nu}
+\bar{G}_{4} \bar{R} 
+\bar{G}_{4,\bar{X}}\left\{ (\bar{\nabla}_{\mu} \bar{A}^{\mu})^{2} 
-\bar{\nabla}_{\mu} \bar{A}_{\nu} \bar{\nabla}^{\nu} \bar{A}^{\mu}\right\} 
\right]\,,
\label{actbare}
\ee
where a bar represents quantities associated with the metric 
tensor $\bar{g}_{\mu \nu}$. For the coupling $\bar{G}_4=\Mpl^2/2+\bar{X}/4$, 
there exists the charged stealth Schwarzschild solution \cite{Chagoya},  
which was shown to be unstable against odd-parity 
perturbations \cite{KMTZ}. 
In this Appendix, we consider the theory related to \eqref{actbare} 
via the disformal transformation:
\be
\bar{g}_{\mu\nu}=g_{\mu\nu}+\Gamma(X)A_{\mu}A_{\nu}\,, 
\label{disformal}
\ee
with $\bar{A}_{\mu}=A_{\mu}$. 
The quantities without a bar represent those associated with 
the metric $g_{\mu \nu}$.
By using Eqs.~(A.21a)-(A.21f) of Ref.~\cite{KNY16}, 
the specific GP theory (\ref{actbare}) is transformed to the action (\ref{action}) 
{\it with} new interactions of the forms 
\be
\frac{1}{4}\alpha_6(X)F^{\mu\nu}F_{\mu\nu}\,,\quad {\rm and} \quad 
\frac{1}{4}\alpha_7(X)A^{\mu}A^{\nu}F_{\mu\rho}{F_{\nu}}^{\rho}\,,
\label{al67}
\ee
where the functions $\alpha_6(X)$ and $\alpha_7(X)$ 
are related to $G_4(X)$ and $\Gamma(X)$.
Thus, the GP theory (\ref{actbare}) cannot be mapped to the BGP theory (\ref{action}) itself via the disformal transformation. 
It is worthy of mentioning that these new interactions 
do not arise in shift-symmetric Horndeski theories, since the term $F_{\mu\nu}$ identically vanishes by taking the scalar limit $A_{\mu} \to \partial_{\mu}\phi$. 

\section*{Acknowledgements}

We thank Ying-li Zhang for useful discussions.
RK is supported by the Grant-in-Aid for Young 
Scientists B of the JSPS No.\,17K14297. 
MM is supported by FCT-Portugal through 
Grant No.\ SFRH/BPD/88299/2012. 
ST is supported by the Grant-in-Aid for Scientific Research 
Fund of the JSPS No.~16K05359 and 
MEXT KAKENHI Grant-in-Aid for 
Scientific Research on Innovative Areas 
``Cosmic Acceleration'' (No.\,15H05890).
MM thanks Tokyo University of Science
for kind hospitality at the early stage 
of this work. RK and ST appreciate CENTRA 
for warm hospitality at the final stage 
of this work.


\end{document}